\documentclass[preprint, prd, nofootinbib, aps, showpacs, showkeys, preprintnumbers]{revtex4-1}
\usepackage{amsmath,amsfonts,amssymb,amsthm,amstext,amscd,eucal}
\usepackage[all]{xy}
\usepackage{epsfig}
\usepackage{amsmath}
\usepackage{amssymb}
\usepackage{graphicx}
\makeatletter \@addtoreset{equation}{section}

\newcommand{\be}{\begin{equation}}
\newcommand{\ee}{\end{equation}}
\newcommand{\bea}{\begin{eqnarray}}
\newcommand{\eea}{\end{eqnarray}}
\newcommand{\fixme}[1]{\textbf{FIXME: }$\langle$\textit{#1}$\rangle$}

\newcommand{\bse}{\begin{subequations}}
\newcommand{\ese}{\end{subequations}}
\newcommand{\bi}{\begin{itemize}}
\newcommand{\ei}{\end{itemize}}

\newcommand{\mpl}{M_{\rm pl}}

\newcommand{\n}{\nonumber\\}

\begin{document}

\preprint{
KUNS-2365\cr
IPM/P-2011/038\cr
\cr\,\,\cr\,\,
\vspace{2cm}}
\title{Gauge-flation and Cosmic No-Hair Conjecture  }

\author{A.~Maleknejad}
\email{azade@ipm.ir}\affiliation {Department of Physics,
Alzahra University
P. O. Box 19938, Tehran 91167, Iran, and\\
School of Physics, Institute for research in fundamental sciences
(IPM), P.O.Box 19395-5531, Tehran, Iran }

\author{M.~M.~Sheikh-Jabbari}
\email{jabbari@theory.ipm.ac.ir} \affiliation{School of Physics,
Institute for Research in Fundamental Sciences (IPM), P. O. Box
19395-5531, Tehran, Iran}

\author{Jiro Soda}
\email{jiro@tap.scphys.kyoto-u.ac.jp }

\affiliation{Department of Physics, Kyoto University, Kyoto, 606-8502, Japan}
\date{\today}

\begin{abstract}
Gauge-flation, inflation from non-Abelian gauge fields, was introduced in \cite{gauge-flation1,gauge-flation2}. In this work, we study the cosmic no-hair conjecture in gauge-flation. Starting from Bianchi-type I cosmology and through analytic and numeric studies we demonstrate that the isotropic FLRW inflation is an attractor of the dynamics of the theory and that the anisotropies are damped within a few e-folds, in accord with the cosmic no-hair conjecture.

\end{abstract}


\maketitle
\section{Introduction}

Inflationary paradigm besides remarkable successes in describing the cosmological data and CMB temperature anisotropies \cite{WMAP7-data}, has a very appealing theoretical and model building feature: inflation considerably relaxes the dependence of late time physics on the pre-inflation initial conditions \cite{Inflation-texts}. A particular set of initial pre-inflation conditions which will be of our interest in this work is the anisotropic but homogeneous initial conditions for metric, the Bianchi type cosmology \cite{Ellis:1998ct}.

Although not directly related to inflationary models,  Wald proved his elegant cosmic no hair theorem \cite{Wald theorem}: In the presence of a cosmological constant, anisotropic but homogenous deviations from de Sitter inflating background with energy momentum tensor respecting strong and dominant energy conditions will be exponentially damped by the dynamics of the system within a few Hubble times scale.  Wald's cosmic no hair theorem has been the basis of the arguments on how inflation washes away the anisotropies, regardless of the details of the inflationary model in question. This expectation is named the cosmic no-hair conjecture.

Inflationary models are generically using scalar fields as inflaton(s) and, dealing  with scalars it is straightforward to check that cosmic no hair conjecture is valid in these models. (For an early confirmation of this conjecture
in the context of chaotic inflationary models see \cite{Sahni}.) This may be seen from the fact that the amplitude of vector perturbations in standard cosmic perturbation theory are (exponentially) damped by the inflationary expansion of the background \cite{Inflation-texts}.

Presence of vector fields, however, may change this picture. This may happen, for example, in the vector inflation models, where there are vector fields turned on in the background as inflaton \cite{Ford:1989me,vector-inflation}.\footnote{We note that vector inflation models, due to broken gauge invariance, generically have ghost problem, inducing instability in the background inflationary dynamics \cite{gauge-flation2,vector-inflation-loophole}.}
Alternatively, one may add U(1) gauge fields with a specifically tailored kinetic term to a standard scalar driven inflation model. The kinetic term can be appropriately chosen such that it can compensate the exponential damping of the vector modes \cite{Jiro-1}. In this latter class of models the modified Maxwell kinetic term provides the setting for violation of assumptions of Wald's cosmic no hair theorem, allowing for growth of the anisotropies. Thus, there exists a counter example to the cosmic no-hair conjecture. Nonetheless, demanding a successful inflationary model restricts the amplitude of anisotropic perturbations to remain small \cite{Jiro-1,Jiro-nature}, compatible with the current observational bounds \cite{Jiro-observational-imprint}. (Anisotropic inflationary models have also been discussed in \cite{anisotropic-others,Jiro-power-law,Jiro-nonAbelian}.)

In \cite{gauge-flation1},  two of us introduced a novel inflationary scenario,
\emph{non-abelian gauge field inflation} or \emph{gauge-flation} for short. In this model, which will be briefly reviewed in section \ref{Section-II}, inflation is driven by non-Abelian gauge field minimally coupled to gravity.\footnote{The gauge-flation  respects the gauge and diffeomorphism invariance and hence is free of the ghost problem of vector inflation models.} It was shown  that non-Abelian gauge field theory can provide the setting for constructing an isotropic and homogeneous inflationary background \cite{gauge-flation2}. Despite of turning on background gauge fields, the isotropy of the background was reinstated  by using the global part of the gauge symmetry of the problem and identifying the $SU(2)$ subgroup of that with the rotation group. We argued that this can be done for \emph{any} non-Abelian gauge group, as any such group has an $SU(2)$ subgroup. Therefore, our discussions can open a new venue for building inflationary models, closer to particle physics high energy models, where non-Abelian gauge theories have a ubiquitous appearance.

Due to the gauge-vector nature of our inflaton field a question that may arise naturally is the classical stability of gauge-flation against the initial anisotropies.
In fact, the rotation symmetry of the universe was essential to set a consistent ansatz with an isotropic universe. Hence, it is not obvious if the special configuration adopted in the previous work is stable against \emph{nonlinear} perturbations in the anisotropic background universe. (Stability against linear perturbations around FLRW gauge-flation background  was shown in \cite{gauge-flation2} in the context of cosmic perturbation theory.) Moreover, one can ask if gauge-flation model is conformed to the
cosmic-no-hair conjecture, and if specific observational features related to anisotropy originated from the gauge fields can be observed.
In this paper, we will tackle these questions. Explicitly, we show that starting with generic anisotropic initial conditions within Bianchi type-I cosmology, dynamics of gauge-flation suppresses the anisotropies. That is, isotropic FLRW cosmology is an attractor of the gauge-flation dynamics.\footnote{Presence and effects of non-Abelian gauge fields in the context of inflationary models have also been studied in \cite{Jiro-nonAbelian} and \cite{non-abelian-others}. These are different than the gauge-flation case because there is also a scalar (inflaton) field in these models. Moreover, the model discussed \cite{Jiro-nonAbelian} involves an action with $SU(2)$ local gauge symmetry, while the one in \cite{non-abelian-others} involves an action with $SU(2)$ global symmetry.}
Since the  anisotropies are damped and basically washed away in the first few e-folds of the inflationary period, gauge-flation predicts a very small
(negligible) statistical anisotropy in the power spectrum of CMB fluctuations.

This paper is organized as follows. In section \ref{Section-II}, we introduce the  gauge-flation setup, and
briefly review the model for isotropic FLRW cosmology and inflation in this model. In section \ref{Section-III},
we consider a Bianchi type-I background metric and, using quasi-de Sitter approximations we study the background anisotropic inflation in gauge-flation setup. Probing space of solutions and  classical trajectories of the model, we show that  regardless of the initial value of anisotropies, our initially anisotropic inflation evolves toward isotropic solution within a few e-folds and after that it effectively mimics the behavior of the isotropic quasi-de Sitter inflation. That is, the isotropic background is the attractor of the system with generic anisotropic initial conditions.
We backup the analytical results of this section by numerical analysis over a range of initial-values. Finally, in section \ref{Section-conclusion}, we summarize our results and make concluding remarks.

\section{Gauge-flation setup }\label{Section-II}

Gauge-flation is based on a non-Abelian gauge theory minimally coupled to Einstein gravity. In \cite{gauge-flation1,gauge-flation2} it was shown that Yang-Mills action for gauge theory part cannot lead to accelerated expansion and one should consider addition of other terms. One particularly convenient choice of the action is
\be\label{The-model}%
S=\int
d^4x\sqrt{-{g}}\left[-\frac{R}{2}-\frac{1}{4}F^a_{~\mu\nu}F_a^{~\mu\nu}+\frac{\kappa}{384
}
(\epsilon^{\mu\nu\lambda\sigma}F^a_{~\mu\nu}F^a_{~\lambda\sigma})^2\right]\,,
\ee
where we have set $8\pi G\equiv \mpl^{-2}=1$ and
$\epsilon^{\mu\nu\lambda\sigma}$ is the totally antisymmetric
tensor. The gauge field strength $F^a_{~\mu\nu}$ is given by
\be\label{F-definition}%
F^a_{~\mu\nu}=\partial_\mu A^a_{~\nu}-\partial_\nu A^a_{~\mu}-g\epsilon^a_{~bc}A^b_{~\mu}A^c_{~\nu}\ ,
\ee
where  $A^a_{~\mu}$, $\mu=0,1,2,3$ label the spacetime indices and $a$ the internal gauge indices. We choose the gauge group to be $SU(2)$ and hence $a=1,2,3$.

Since $\sqrt{-g}\epsilon^{\mu\nu\lambda\rho}$ is independent of the metric,
the metric dependence of the $\kappa F^4$ term appears only through $\det g$.
 Its contribution to the
energy momentum tensor, denoted by $T^{\kappa}_{\mu\nu}$, is of the form
\be
T^\kappa_{\mu\nu}=-\rho_\kappa(t) g_{\mu\nu}\,.
\ee
In the absence of the Yang-Mills term, equation of motion for the gauge field gives a constant $\rho_\kappa$ and hence, if $\kappa >0$, we have a de Sitter expansion phase. Addition of the Yang-Mills term, however, changes this behavior and eventually ends inflation. (Note that energy momentum of the Yang-Mills term $T_{\mu\nu}^{YM}$ is traceless and hence cannot sustain accelerated expansion phase.)

To be more precise, defining the isotropic expansion rate $H$ as
$$\sqrt{-g}|_{\Sigma_t}=e^{3\alpha},\quad H\equiv\dot\alpha,$$
where $\Sigma_t$ are the space-like constant comoving time hypersurfaces, we can define
the parameter
\be\label{slow-roll-def}
\epsilon=-\frac{\dot H}{H^2}  \,,
\ee
which is nothing but the slow-roll parameter in the standard slow-roll inflationary scenario. This parameter describes to what extent the universe is close to de Sitter spacetime. In this paper we will use the expression ``quasi-de Sitter'' inflation for denoting a background with $\epsilon\ll 1$.
Therefore, quasi-de Sitter  captures the standard isotropic FLRW slow-roll inflation as well as allowing for anisotropic cases.

From the above argument, demanding quasi-de Sitter inflation ($\epsilon\ll1$) is equivalent to the condition $\frac{\rho_{_{YM}}}{\rho_\kappa}\ll1$ (where $\rho_{YM}$ is $T_{\mu\nu}^{YM}$ projected on the vector normal to $\Sigma_t$).
Namely, when the $F^4$ contributions to the energy density $\rho_{\kappa}$ dominate over the Yang-Mills contributions $\rho_{_{YM}}$, the universe is quasi-de Sitter. The time evolution of the system then slowly increases
$\rho_{_{YM}}$ with respect to $\rho_{\kappa}$, and when $\rho_{_{YM}}\sim \rho_{\kappa}$, the quasi-de Sitter inflation ends.

The above analysis can be made more explicit if we consider an isotropic homogenous background. This was the case studied in some detail \cite{gauge-flation2}. Here, we review some of the calculations of \cite{gauge-flation2} as a warmup and also for fixing some conventions and notations we will use in the anisotropic case of the next section. We start with the background FLRW metric
\be \label{FLRW}
ds^2=-dt^2+a^2(t)\delta_{ij}dx^idx^j\,.
\ee
Choosing the temporal gauge $A^a_{~0}=0$, we can set the ansatz
\bea  \label{A-ansatz-background}
A^a_{~i}=\psi(t)e^a_{~i}=a(t)\psi(t)\delta^a_i , %
\eea%
where $e^i_a$ are the triads on constant $t$ slices, is consistent with the dynamics and identifies the combination of the gauge fields for which the rotation symmetry violation caused by turning on vector (gauge) fields in the background is compensated  (or undone)  by the gauge transformations, leaving us with a rotationally invariant background. Note that under both of 3d diffeomorphisms and gauge transformations $\psi(t)$ behaves as a
genuine scaler field, nonetheless it appears that equations take a simpler form once expressed in terms of $\phi$
\be\label{phi}
\phi(t)=a(t)\psi(t)\,,
\ee
or equivalently $A^a_{~i}=\phi(t)\delta^a_{~i}$.

Evaluating \eqref{The-model} for the FLRW metric and the ansatz \eqref{A-ansatz-background}, we obtain
\be
\mathcal{L}_{red}=\frac{3}{2}(\frac{\dot{\phi}^2}{a^2}-\frac{g^2\phi^4}{a^4}+\kappa
\frac{g^2\phi^4\dot{\phi}^2}{a^6})\,, \ee
which is the Lagrangian governing dynamics of the system in the homogenous-isotropic sector \cite{gauge-flation2}.
Using $\mathcal{L}_{red}$ one can compute the energy density $\rho$
and pressure $P$
\be%
\rho=\rho_{_{YM}}+\rho_{\kappa}\,,\qquad
P=\frac13\rho_{_{YM}}-\rho_\kappa\,,
\ee%
where $\rho_{YM}$ and $\rho_\kappa$ are respectively contributions of  Yang-Mills and $\kappa$ terms of the action
\be\label{rho0-rho1}%
\rho_{_{YM}}=\frac{3}{2}(\frac{\dot{\phi}^2}{a^2}+\frac{g^2\phi^4}{a^4})\,,\qquad
\rho_{\kappa}=\frac32\frac{\kappa g^2\phi^4\dot{\phi}^2}{a^6}\,.
\ee%
Note that in order to have positive energy we consider positive $\kappa$ values,
hence $\rho_{_{YM}}$ and $\rho_\kappa$ are both positive quantities.\footnote{Note also that the non-Abelian character of the fields, besides in resolution of anisotropy of the background caused by turning on the gauge fields, also shows up in the fact that the $\kappa$-term gives a nonvanishing contribution to the energy density $\rho$, which is essential for having inflation.}

The Einstein equations, the Friedmann equation and the evolution equation for $H$,   are then obtained as%
\bea \label{cosm1}
&~&H^2=\frac{1}{2}(\frac{\dot{\phi}^2}{a^2}+\frac{g^2\phi^4}{a^4}+\kappa \frac{g^2\phi^4\dot{\phi}^2}{a^6})
,\qquad
\dot{H}=-(\frac{\dot{\phi}^2}{a^2}+\frac{g^2\phi^4}{a^4})  \,.
\eea
Using the Friedmann equations \eqref{cosm1} and definitions \eqref{rho0-rho1} and \eqref{slow-roll-def}, we have
\be\label{epsilon-rho0-rho1}
\epsilon= \frac{2\rho_{_{YM}}}{\rho_{_{YM}}+\rho_\kappa}\,.
\ee%
Here, we can see that in order to have quasi-de Sitter inflation the $\kappa$-term contribution
$\rho_\kappa$ should dominate over the Yang-Mills contributions
$\rho_{_{YM}}$, i.e. $\rho_\kappa\gg \rho_{_{YM}}$.

The Yang-Mills equations reduces to a single equation for$\phi$ field:
\bea
\label{phi-eom}
(1+\kappa\frac{g^2\phi^4}{a^4})\frac{\ddot{\phi}}{a}+(1+\kappa\frac{\dot{\phi}^2}{a^2})\frac{2g^2\phi^3}{a^3}+
(1-3\kappa\frac{g^2\phi^4}{a^4})\frac{H\dot{\phi}}{a}=0 \,.
\eea
In the $\kappa$ dominant limit, i.e. when  $\epsilon\ll 1$, the above equation becomes
\bea
\label{phi-limit}
\frac{d}{dt} \left( \frac{\phi^2 \dot{\phi}}{a^3} \right) \simeq 0 \,,
\eea
where $\simeq$ means equality to the leading order in the parameter $\epsilon$.
This implies $\rho_\kappa \simeq {\rm constant}$ and then $H \simeq {\rm constant}$.
In this limit,
by integrating Eq.(\ref{phi-limit}), we also have a relation
$\phi \propto \exp Ht \propto a $. Hence, $\psi$ must be almost constant
during quasi-de Sitter inflation. Since we also have a relation $\dot{\phi}\simeq H\phi$,
  we obtain the formula
\bea \label{epsilon-x}
\epsilon\simeq\psi^2(\gamma+1) ,  %
\eea%
where $\gamma$ is defined as
\be\label{x-def}
\gamma\equiv\frac{g^2\psi^2}{H^2}\,.
\ee%
Thus, to have a sufficient $e$-folding number of inflation, we need sub-Planckian field values,
$\psi \ll 1$. (Note that we are working in $M_{pl}=1$ units.)

To summarize,  assuming $\rho_\kappa\gg\rho_{_{YM}}$, we have a successful isotropic  quasi-de Sitter inflation period, during which
$H$ and $\psi$ are almost constant. Since $\epsilon\propto\psi^2$, the quasi-de Sitter condition demands sub-Planckian field values.
The time evolution will then increase $\rho_{_{YM}}$ with respect to
$\rho_\kappa$, and when $\rho_{_{YM}}\sim \rho_\kappa$ the
slow-roll inflation ends. Noting that
$\rho+3P=2(\rho_{_{YM}}-\rho_\kappa)$, inflation (accelerated
expansion phase) ends when $\rho_{_{YM}}>\rho_\kappa$.

\section{Gauge-fation and Cosmic No-Hair}\label{Section-III}

In this section we  study  gauge-flation  in a homogenous but anisotropic background. In this way we examine the generality of the isotropic FLRW gauge-flation.  Here, for practical reasons we consider gauge-flation in an axially symmetric Bianchi type-I setup but we believe that our results are valid for more general anisotropic cases.

Bianchi type-I axially symmetric metric is described by the line element
\be \label{axi-metric}
ds^2=-N^2dt^2+e^{2\alpha(t)}\big(e^{-4\sigma(t)}
dx^2+e^{2\sigma(t)}(dy^2+dz^2)\big),
 \ee
where $N$ is the lapse function, $e^{\sigma(t)}$ represents the anisotropy and, $e^{\alpha(t)}$ is the isotropic scale factor.
Given the symmetries of the metric, as before we choose the temporal gauge for the gauge fields
$A^a_{~0}=0$, and make the following modification to the isotropic ansatz \eqref{A-ansatz-background}
\bea\label{axi-ansatz}
A^a_{~i}=e^a_{~i}\psi_i,\qquad \psi_1\equiv\frac{\psi}{\lambda^2},\qquad \psi_2=\psi_3\equiv\lambda\psi,
\eea
where $\psi_i$ act as two scalar fields and $e^a_{~i}$ are the triads associated with spatial metric.
The gauge field one-form of the above ansatz is hence of the form  \footnote{A similar ansatz has been considered in \cite{Jiro-nonAbelian}.}
\be \label{ansatz1}
A=A^a_{~i}T^adx^i=e^{\alpha(t)-2\sigma(t)}\frac{\psi(t)}{\lambda^2(t)} T^1dx+
e^{\alpha(t)+\sigma(t)}\lambda(t)\psi(t)(T^2dy+T^3dz)\,,
\ee
where $T^a$ are the $SU(2)$ gauge group generators,
$ [T_a,T_b]=i\epsilon^c_{~ab}T_c$. It turns out that, similar to  the isotropic case \eqref{phi}, the equations take a simpler form once written in terms of $\phi$
\bea\label{field}
\phi(t)=a(t)\psi(t)\,,
\eea
where
\be
a(t)\equiv e^\alpha\,,
 \ee
is the isotropic scale factor.

Inserting the axi-symmetric Bianchi metric and the gauge field ansatz into the gauge-flation action \eqref{The-model}, the total reduced (effective) action is obtained as
\bea
S_{red}&=&\int dt \frac{a^3}{N}\biggl[-3\dot \alpha^2+\dot\sigma^2\big(3+\frac{(2+\lambda^6)}{\lambda^4}\frac{\phi^2}{a^2}\big)
+\dot\sigma\frac{\big(\lambda^{-4}(\lambda^6-1)\phi^2\dot{\big)}}{a^2}+\frac{(1+2\lambda^6)}{2\lambda^4}\frac{\dot\phi^2}{a^2}\n
&+&2\frac{(\lambda^6-1)}{\lambda^4}\frac{\dot\lambda}{\lambda}\frac{\dot\phi\phi}{a^2}+\frac{(2+\lambda^6)}{\lambda^4}\frac{\dot\lambda^2}{\lambda^2}\frac{\phi^2}{a^2}
-N^2\frac{(2+\lambda^6)}{2\lambda^2}\frac{g^2\phi^4}{a^4}+
\frac{3}{2}\frac{\kappa g^2\phi^4}{a^4}\frac{\dot\phi^2}{a^2}
\biggr]\,,
\eea
where a dot denotes derivative with respect to the time coordinate $t$. As we see, the above action depends only on $\dot\sigma$ and not $\sigma$.  Therefore, momentum conjugate to $\sigma$ is a constant of motion. This constant may be chosen on the physical basis that the anisotropy $\dot\sigma$ should vanish for the isotropic gauge field, i.e. when $\lambda^2=1$. With this initial condition we obtain
\be\label{const1}
\dot\sigma=-\frac{\big(\lambda^{-4}(\lambda^6-1)\phi^2\dot{\big)}}{2a^2\big(3+\lambda^{-4}(2+\lambda^6)\frac{\phi^2}{a^2}\big)}\,.
\ee

We can deduce the field equations corresponding to $\phi$, $\lambda$ and $\dot\alpha$ from the above effective action. However, instead, we find it  more convenient to determine the energy-momentum tensor and study the Einstein equations, as well as the field equation for $\lambda$. Substituting the ansatz \eqref{ansatz1} and the metric \eqref{axi-metric} into $T_{\mu\nu}$ for the gauge fields, we obtain a diagonal homogenous tensor, $$T^{\nu}_{~\mu}=\textmd{diag}(-\rho(t),P_1(t),P_2(t),P_2(t)).$$ One can decompose the energy density $\rho$ as
\be
\rho=\rho_\kappa+\rho_{_{YM}},
\ee
where $\rho_\kappa$ and $\rho_{_{YM}}$  are respectively  contributions of $F^4$ and Yang-Mills terms given by
\bea\label{k-rho}
\rho_\kappa&=&\frac{3}{2}\frac{\kappa g^2\phi^4}{a^4}\frac{\dot{\phi}^2}{a^2},\\
\label{ym-rho}
\rho_{_{YM}}&=&\frac{3}{2}\left[\frac{1}{3\lambda^4}\big(\frac{\dot\phi}{a}-2(\dot\sigma+\frac{\dot\lambda}{\lambda})\frac{\phi}{a}\big)^2
+\frac{2\lambda^2}{3}\big(\frac{\dot\phi}{a}+(\dot\sigma-\frac{\dot\lambda}{\lambda})\frac{\phi}{a}\big)^2
+\frac{(2+\lambda^6)}{3\lambda^2}\frac{g^2\phi^4}{a^4}\right].
\eea
Note that $\rho_\kappa$ is only a function of $\phi$ and not $\lambda$.
As mentioned before, we consider positive $\kappa$ values, so $\rho_\kappa$ and $\rho_{_{YM}}$ are both positive quantities.
Furthermore, $P_1$ and $P_2$ are
\bea\label{axi-P}
P_1&=&P-\frac23\tilde P,\\ \label{axi-tildeP}
P_2&=&P+\frac13\tilde P,
\eea
where the isotropic pressure $P$ and the anisotropic pressure $\tilde P$ are given by
\bea
P&=&-\rho_\kappa+\frac{1}{3}\rho_{_{YM}}\\ \nonumber
\tilde P&=&(1-\lambda^6)\big[\frac{1}{\lambda^4}(\frac{\dot\phi}{a}-2(\dot\sigma+\frac{\dot\lambda}{\lambda})\frac{\phi}{a})^2-\frac{1}{\lambda^2}\frac{g^2\phi^4}{a^4}\big]
-3\lambda^2(\frac{\dot\lambda}{\lambda}+\dot\sigma)\big[2\frac{\dot\phi}{a}-(\dot\sigma+\frac{\dot\lambda}{\lambda})\frac{\phi}{a}\big]\frac{\phi}{a}.
\eea
As we see, in the isotropic case  $\lambda^2=1$ ($\dot{\sigma}=0$ and $\dot\lambda=0$), $\tilde P$ vanishes and hence $P_1=P_2=P$.

The independent gravitational field equations are
 \bea
\label{lambda}
&~&\dot{\alpha}^2-\dot{\sigma}^2=\frac{\rho}{3},\\
\label{dsigma}
&~&\ddot{\sigma}+3\dot{\alpha}\dot{\sigma}=\frac{P_2-P_1}{3}=\frac{\tilde P}{3},\\
\label{xi} &~&\ddot{\alpha}+3\dot{\sigma}^2=-\frac{3\rho+P_1+2
P_2}{6}=-\frac23\rho_{_{YM}}.
\eea
Note that \eqref{dsigma} implies that the evolution of $\dot\sigma$ depends on the anisotropic part of the pressure $\tilde P$, i.e. $\tilde P$ is the source for the anisotropy $\dot\sigma$ and in the absence $\tilde P$ the anisotropy $\dot\sigma$ is exponentially damped, with time scale $\dot\alpha^{-1}$.

\subsection{Analysis in Quasi-de Sitter regime}

So far we computed $T_{\mu\nu}$, the reduced action and $\dot\sigma$ without assuming the
$\kappa$-term dominance.
At this point, we simplify and analyze the equations assuming quasi-de Sitter inflation, in the sense that the parameter $\epsilon$
$$\epsilon=-\frac{\ddot{\alpha}}{\dot\alpha^2} $$
is  small during  inflation.
Note that we only impose the quasi-de Sitter condition on \emph{the isotropic parts of the expansion} and the rest of variables are \emph{not} enforced to have slow dynamics.

From the combination of \eqref{lambda} and \eqref{xi} we obtain $\epsilon$ in terms of $\rho_\kappa$, $\rho_{_{YM}}$ and $\dot\sigma$
\be
\epsilon=\frac{2\rho_{_{YM}}+9\dot\sigma^2}{\rho_\kappa+\rho_{_{YM}}+3\dot\sigma^2}\,,
\ee
which implies $\rho_{\kappa}\gg\rho_{_{YM}}$ and $\rho_{\kappa}\gg\dot\sigma^2$
for having a very small $\epsilon$. Therefore, from \eqref{lambda} and after using \eqref{k-rho}, we obtain
\be
\rho_\kappa=\frac{3}{2}\frac{\kappa g^2\phi^4}{a^4}\frac{\dot\phi^2}{a^2}=3\dot\alpha^2+O(\epsilon)\,.
\ee
As we have learned in the previous section, the constancy of $\rho_\kappa$ implies
\be\label{aprodpsi}
\dot\phi\simeq\dot\alpha\phi\,.
\ee
Moreover, noting that $\tilde P$ is a term from the contribution of Yang-Mills part, and the corresponding isotropic pressure is also positive,  we learn that $\rho_{\kappa}\gg \tilde P$.
Then, using \eqref{dsigma} we obtain $\dot\alpha\dot\sigma\sim \tilde P$ which gives
\be
\dot\sigma^2\sim\frac{\tilde P^2}{\rho_\kappa}\ll \tilde P\,,
\ee
and as $\tilde P$ is a quantity at most of same order as $\rho_{_{YM}}$, we obtain $\dot\sigma^2\ll\rho_{_{YM}}$.
As a result, during quasi-de Sitter inflation, we have the following approximation for $\epsilon$
\be\label{aproxepsilon}
\epsilon\simeq\frac{2\rho_{_{YM}}}{3\dot\alpha^2}\simeq\frac{2\rho_{_{YM}}}{\rho_\kappa},
\ee
where $\simeq$ means equality to the leading order of $\epsilon$.
Using the above relation and \eqref{aprodpsi}, ignoring the $\dot\sigma$ terms in $\rho_{_{YM}}$ , we find
\bea
\epsilon\simeq\frac{\phi^2}{3a^2\lambda^4}\bigg[(2+\lambda^6)(\lambda^2\gamma+\frac{2\dot\lambda^2}{\dot\alpha^2\lambda^2})-
\frac{4\dot\lambda}{\dot\alpha\lambda}(1+\lambda^6)+(1+2\lambda^6)\bigg]\,,
\eea
where $\gamma\equiv\frac{g^2\phi^2}{a^2\dot\alpha^2}$.
As we will show through analytical calculations, $\frac{\dot\lambda}{\dot\alpha\lambda}$ is at most an order one quantity. Therefore, in order to have a successful quasi-de Sitter inflation ($\epsilon\ll1$), we should have
\be\label{phi-small}
\frac{\lambda\phi}{a}\ll1 \ ,   \qquad  \frac{\phi}{a\lambda^2}\ll1\,,
\ee
for large and small $\lambda$, respectively.
In other words, recalling \eqref{axi-ansatz}, similar to the isotropic inflation,  our field values $\psi_i$'s should have (physically
reasonable) sub-Planckian values during the quasi-de Sitter inflation.

From $\dot\alpha\dot\sigma\sim \tilde P$ and $\tilde P < \rho_{YM}$, we obtain
\be\label{uppersigma}
\frac{\dot\sigma^2}{\dot\alpha^2} \sim \left(\frac{\tilde P}{\rho_{\kappa}}\right)^2
        < \left(\frac{\rho_{YM}}{\rho_{\kappa}}\right)^2 \sim \epsilon^2 \,.
\ee
That is,  that the value of $\frac{\dot\sigma^2}{\dot\alpha^2}$ in our anisotropic gauge-flation model cannot be more than $\epsilon^2$.

To summarize, so far we have shown that during the quasi-de Sitter inflation still means $\rho_\kappa$ dominance. Precisely, $\epsilon\simeq\frac{2\rho_{_{YM}}}{\rho_\kappa}$ and
$\frac{\dot\sigma^2}{\dot\alpha^2}$, which represents the
 anisotropy of the expansion of the universe, can be at most of
order $\epsilon^2$. To proceed further we need to analyze dynamical field equations.

Using \eqref{lambda}, \eqref{aproxepsilon} and ignoring $\dot\sigma$ terms, we obtain the following relation to the leading order of $\epsilon$,
\be
3\dot\alpha^2 \simeq\rho_\kappa \,,\qquad
\frac{\phi}{a} \simeq\big(\frac{2}{\kappa g^2}\big)^{\frac16}.
\ee
Neglecting $\dot\sigma$ terms, we can deduce the field equation corresponding to $\lambda$ as
\bea\label{lambdaeq}
\bigg((2+\lambda^6)(\lambda\ddot\lambda+\dot\alpha\lambda\dot\lambda+2\frac{\dot\phi}{\phi}\lambda\dot\lambda)-6\dot\lambda^2\bigg)
\frac{\phi^2}{a^2}+\lambda^2(\lambda^6-1)(\frac{\phi\ddot\phi}{a^2}+\frac{\dot\alpha\phi\dot\phi}{a^2}+\frac{\lambda^2g^2\phi^4}{a^4})=0,\quad
\eea
which, using \eqref{aprodpsi} and keeping the leading orders, can be simplified to
\be\label{eqlambda1}
(2+\lambda^6)(\lambda\ddot\lambda+3\dot\alpha\lambda\dot\lambda)-6\dot\lambda^2
+\lambda^2(\lambda^6-1)(2+\lambda^2\gamma)\dot\alpha^2\simeq0.\quad
\ee
One can see that $\lambda=0$ is a singular point of the above equation and that the dynamics do not mix $\lambda>0$ and $\lambda<0$ regions. That is, if $\lambda$ is positive (negative) initially, it always remains positive (negative) during the inflation.

Since \eqref{eqlambda1} is a nonlinear second order differential equation which has no explicit time dependence, its solution will be of the form $\dot\lambda=\dot\lambda(\lambda)$ and hence
\be
\ddot\lambda=\frac{d\dot\lambda}{d\lambda}\dot\lambda\,.
\ee
In terms of derivatives with respect to $dN=\dot\alpha dt $, and denoting $\frac{d}{dN}$ by a prime, we obtain
\be
\frac{d\lambda'}{d\lambda}\simeq-3+\frac{1}{(2+\lambda^6)}\big(\frac{6\lambda'}{\lambda}
-\frac{\lambda(\lambda^6-1)}{\lambda'}(2+\gamma\lambda^2)\big),
\ee
 which implies that $\lambda'(\lambda)$ is an odd function of $\lambda$, $\lambda'(-\lambda)=-\lambda'(\lambda)$.

\begin{figure}[h]
\includegraphics[angle=0, width=80mm, height=70mm]{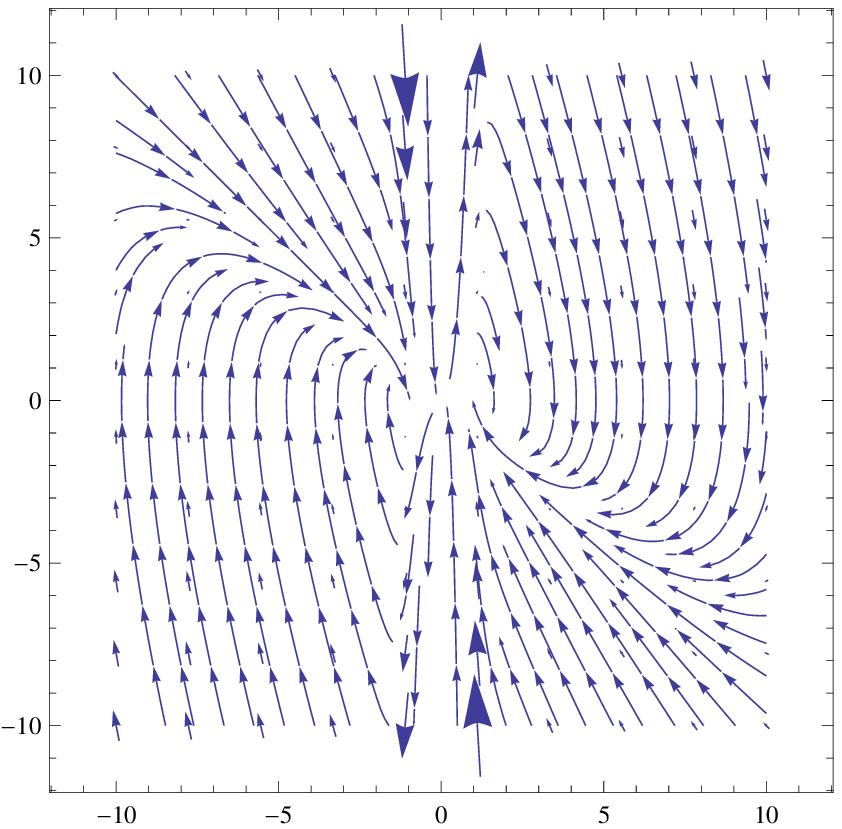}
\includegraphics[angle=0, width=80mm, height=70mm]{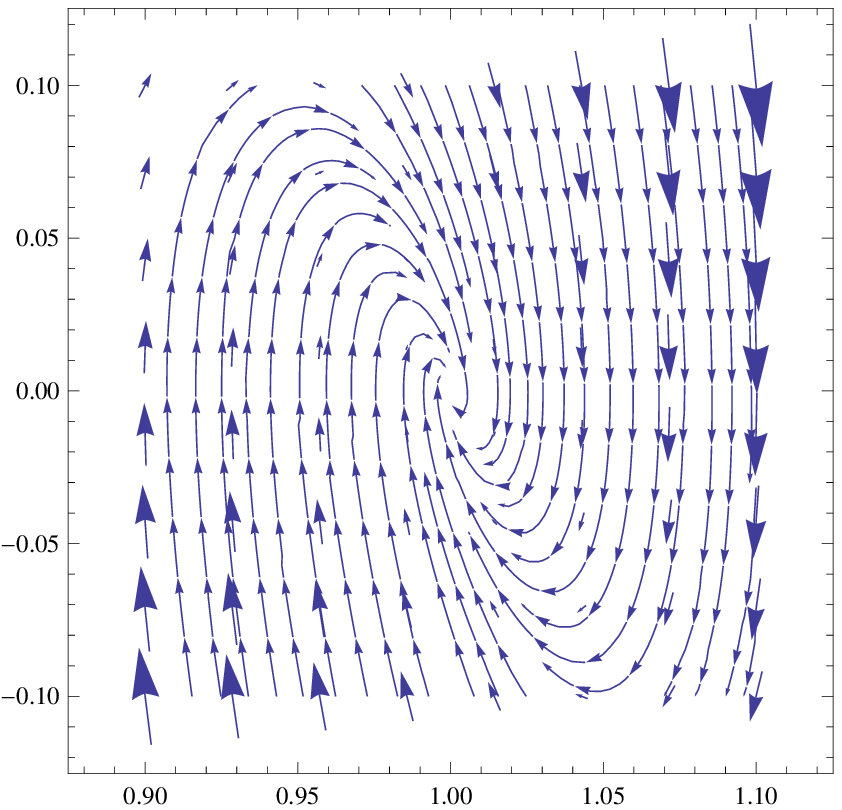}
\caption{The  phase diagram in the $\lambda'-\lambda$ plane, the vertical axis is $\lambda'$ and the horizontal is $\lambda$. Both figures show existence of attractor at $|\lambda|=1$, corresponding to the isotropic FLRW background. The left figure shows the phase diagram over a large range of values for $\lambda$, while  the right figure shows the phase diagram for $\lambda$ in the vicinity of the attractor solution $\lambda=1$. The left figure explicitly exhibits the $\lambda'(\lambda)=-\lambda'(-\lambda)$ symmetry.
}\label{phaseDiag}
\end{figure}

 Using \textsf{Mathematica}, the above equation can be studied by the phase diagram method, and in Figure \ref{phaseDiag}, we have presented the behavior of the solutions
in the $\lambda'-\lambda$ plane. Apparently, all of trajectories approach to the isotropic fixed point $\lambda^2 =1$. Next, we give the asymptotic analysis to confirm
that the isotropic inflation is an attractor in the phase space.

\subsection{Asymptotic Analysis}

To study the behavior of the system more precisely, it is convenient to rewrite  \eqref{eqlambda1} as
\be\label{eqlambda}
(2+\lambda^6)(\frac{\lambda''}{\lambda}+3\frac{\lambda'}{\lambda})-6\frac{\lambda'^2}{\lambda^2}
+(\lambda^6-1)(2+\lambda^2\gamma)\simeq0\,,
\ee
in which we distinguish three different regions for the value of $\lambda$:
\begin{itemize}
\item[I)] $~\lambda^6-1$ close to zero ($\lambda^6\simeq1$),
\item[II)] $~\lambda^6$ in the vicinity of zero,
\item[III)] large $\lambda^6$ limit.
\end{itemize}
Furthermore, defining
 \be
 \Sigma\equiv\frac{\dot\sigma}{\dot\alpha}=\sigma'\,,
 \ee
and using \eqref{aprodpsi} in \eqref{const1}, we can write $\Sigma$ as
 \be\label{eqsigma}
 \Sigma\simeq-\frac13\big((\lambda^6+2)\frac{\lambda'}{\lambda}+(\lambda^6-1)\big)\frac{\phi^2}{a^2\lambda^4}\,.
 \ee

We will solve \eqref{eqlambda} in the above three different limits, and by inserting the corresponding solutions in \eqref{eqsigma}, we  study the behavior of $\Sigma$.
\begin{itemize}
 \item[I)]{In this case we can approximate $\lambda$ as%
\be\label{dellambda}
\lambda=\pm1+\frac16\delta\lambda,\quad |\delta\lambda|\ll1\,,
\ee
and equation \eqref{eqlambda} as
 \be
 \delta\lambda''+3\delta\lambda'+2(2+\gamma)\delta\lambda\simeq0\,.
 \ee
Then, solving the above equation, one obtains the following solution for $\lambda$ as a function of number of $e$-folds, $\dot\alpha t$,
\be\label{dellamdasol}
\delta\lambda\simeq 6e^{-\frac32\dot\alpha t}\big(A_1\cos(\sqrt{\frac74+2\gamma}~\dot\alpha t)+A_2\sin(\sqrt{\frac74+2\gamma}~\dot\alpha t)\big).
\ee
As we see, the damping ratio is equal to $\frac{3}{2\sqrt{2(2+\gamma)}}$, which is less than one. Thus, $\delta\lambda$ shows a damped oscillation and is damped within one or two number of $e$-folds.
From the combination of \eqref{dellamdasol} and \eqref{dellambda}, we can determine $\lambda$, which has the following form in the vicinity of $\lambda=\pm1$
\be\label{lambda1}
\lambda\simeq \pm1+e^{-\frac32\dot\alpha t}\big(A_1\cos(\sqrt{\frac74+2\gamma}~\dot\alpha t)+A_2\sin(\sqrt{\frac74+2\gamma}~\dot\alpha t)\big).
\ee
We see that $\lambda$ is exponentially approaching $\pm1$ and the trajectory meets the attractors $\lambda=\pm1$.

Inserting \eqref{lambda1} in \eqref{eqsigma}, we can determine $\Sigma$
\be
\Sigma\simeq-\frac{\phi^2}{3a^2}\big(\frac{1}{2}\frac{\delta\lambda'}{\delta\lambda}+1\big)\delta\lambda,
\ee
which implies that $\Sigma$ and $\delta\lambda$ have opposite signs.

As we see, for $\lambda^2$  in the vicinity of one $\dot\sigma$ has the following behavior
\be
|\frac{\dot\sigma}{\dot\sigma_0}|\leq e^{-\frac32\dot\alpha t},
\ee
here the subscript 0 denotes an initial value.
 Thus $|\dot\sigma|$ is exponentially damped, with a time scale $2/(3\dot\alpha)$.}

 \item[II)] {In the limit of very small $\lambda^6$ values ($|\lambda|\ll1$) and considering the leading orders, equation \eqref{eqlambda} has the following form
 \be
 \frac{\lambda''}{\lambda}+3\frac{\lambda'}{\lambda}-3\frac{\lambda'^2}{\lambda^2}-1\simeq0\,,
 \ee
which can be simplified as
\be\label{lam1}
(\frac{1}{\lambda^2}\big)''+3\big(\frac{1}{\lambda^2}\big)'+2\frac{1}{\lambda^2}\simeq0\,.
\ee
Solving the above equation, we obtain
\be\label{lambda0}
\frac{1}{\lambda^2}\simeq A_1e^{-2\dot\alpha t}+A_2e^{-\dot\alpha t}\,,
\ee
which represents an exponential increase in $|\lambda|$ value with time scale of the order $\dot\alpha^{-1}$.
Thus, in the limit of initially very small $\lambda^2$ values, $|\lambda|$ is growing very rapidly
and escaping quickly from the vicinity of zero. As a result, the above approximate solution is only applicable in first few $e$-folds where $\lambda^2$ is far from one.

Eq. \eqref{lambda0} shows that for all possible values of $A_1$ and $A_2$, $\lambda$ monotonically increases, but interestingly this is not necessarily the case for $\dot\sigma$. In order to investigate this fact more precisely, we determine $\Sigma$ for two different initial conditions in which (i) $A_1=0$ and (ii) $A_2=0$.
\begin{itemize}
\item[(i)]{ Putting $A_1=0$ in \eqref{lambda0}, we have $\lambda=\lambda_0e^{\frac12\dot\alpha t}$, then after inserting in \eqref{eqsigma} we have
    \be
    \Sigma\simeq-\frac{\psi^2}{2}\lambda^2\simeq-\frac{\lambda_0^2\psi^2}{2}e^{\dot\alpha t},
    \ee
i.e. $|\Sigma|$ exponentially increases in time. (Note that $\psi=\frac{\phi}{a}$ is  a constant in the leading order of $\epsilon$.) As mentioned before, due to the exponential growth of $\lambda$, the approximation \eqref{lambda0} is only applicable for the first couple of $e$-folds. Therefore,  the growth of $|\Sigma|$ can happen only during the first few $e$-folds, after that $\lambda$ gets close to one and $|\Sigma|$ is exponentially damped. Note that although the phase of exponential growth of $|\Sigma|$ lasts for the first couple of e-folds. 
}
\item[(ii)]{Putting $A_2=0$, we obtain $\lambda=\lambda_0e^{\dot\alpha t}$, leading to
\be
\Sigma\simeq-\frac{\psi^2}{3\lambda^4}\simeq-\frac{\psi^2}{3\lambda_0^4}e^{-4\dot\alpha t},
\ee
which is quickly damped.
}
\end{itemize}
 Note that in this limit $\lambda^6\ll1$ and $\Sigma$ has always  a negative sign.

}
\item[III)]{ In the limit of  large $\lambda^6$ values ($|\lambda|\gg1$) and recalling \eqref{phi-small}, $\frac{\lambda\phi}{a}\ll1$, we have $\lambda^2\gamma\ll1$. As a result, up to the leading orders, we obtain the following approximation for \eqref{eqlambda}
\be\label{lam2}
\lambda''+3\lambda'+2\lambda\simeq0,
\ee
which is identical to \eqref{lam1} that governs the evolution of $\frac{1}{\lambda^2}$ in the limit of $|\lambda|\ll1$. Thus, the behavior of $\lambda$ in the limit of $|\lambda|\gg1$ is identical to the behavior of $\frac{1}{\lambda^2}$ in the limit of $|\lambda|\ll1$, however, for completeness we present a more detailed analysis of this equation.

Solving the above equation, we obtain the following solution for $|\lambda|\gg1$
\be\label{lambdabig}
\lambda\simeq A_1e^{-2\dot\alpha t}+A_2e^{-\dot\alpha t},
\ee
 which has an exponentially damping behavior.
 Thus, in the limit of initially very large $\lambda^6$ values, $|\lambda|$ is damped very strongly and after one or two number of e-folds $\lambda$ becomes close to one and the approximate equation \eqref{lam2} is not applicable any more.
For all possible values of $A_1$ and $A_2$ in \eqref{lambdabig}, $\lambda$ monotonically decreases, but interestingly this is not necessarily the case for $\dot\sigma$. To study this fact more explicitly we determine $\Sigma$ for two initial conditions in which (i) $A_1=0$ and (ii) $A_2=0$.

\begin{itemize}
\item[(i)]{ Putting $A_1=0$ in \eqref{lambdabig}, we have $\lambda=\lambda_0e^{-\dot\alpha t}$ which after inserting in \eqref{eqsigma} yields
 monotonically increasing $\Sigma$
    \be
    \Sigma\simeq\frac{\psi^2}{\lambda^4}\simeq\frac{\psi^2}{\lambda_0^4}e^{4\dot\alpha t}\,.
    \ee
Note that due to the exponential damping behavior of $\lambda$, the approximation \eqref{lambdabig} is applicable only for at most one or two number of e-folds. Hence, the growth of $\Sigma$ in this case, happens during the first few e-folds and after $\lambda$ gets close to one, $\Sigma$ is exponentially damped.
}
\item[(ii)]{ On the other hand, putting $A_2=0$, we obtain $\lambda=\lambda_0e^{-2\dot\alpha t}$, which gives $\Sigma$ as
\be
\Sigma\simeq\frac{\psi^2\lambda^2}{3}\simeq\frac{\psi^2\lambda_0^2}{3}e^{-4\dot\alpha t},
\ee
which is quickly damped.

In this limit $\lambda^6\gg1$ and the sign of $\Sigma$ is always positive.}
\end{itemize}
}
\end{itemize}

To summarize, assuming a system which undergoes quasi-de Sitter inflation
in the sense that $\epsilon$ is very small, we determined $\lambda$ and $\Sigma$. We see that regardless of the initial $\lambda$
values, all solutions converge to $\lambda^2=1$, within the few first e-folds.
Note that,  $\lambda^2=1$
corresponds to two values $\lambda=\pm1$, which are the isotropic solutions. As we saw before, $\lambda$ can not pass through zero during its evolution, so its sign does not change in time. As a result, as we have shown analytically and will be demonstrated numerically in the next subsection, system's trajectory eventually meets
its attractor solution
$$\lambda\rightarrow1~~~\textmd{if}~~~\lambda_0>0,$$
$$\lambda\rightarrow-1~~ \textmd{if}~~~\lambda_0<0.$$
Furthermore, comparing \eqref{lam1} and \eqref{lam2} which are the approximate forms of \eqref{eqlambda} in the limits of very small and very large $\lambda^6$ values, we find out that the behavior of $\lambda$ in the limit of $|\lambda|\gg1$ is identical to the behavior of $\frac{1}{\lambda^2}$ in the limit of $|\lambda|\ll1$.

Although $\lambda^2$ evolves toward the FLRW isotropic solutions, it is shown that in some solutions, $|\Sigma|$ grows rapidly at the first few e-folds saturating our upper bound of $\Sigma\sim\epsilon$ for a short time. However, this growth  stops fast (within a couple of e-folds) and $\Sigma$ is damped for the rest of the quasi-de Sitter inflation.

\subsection{Numerical Analysis}\label{numerical-analysis-section}

As seen from the action \eqref{The-model}, our gauge-flation model
has two parameters, the gauge coupling $g$ and the coefficient of
the $F^4$ term $\kappa$. On the other hand, the field degrees of
freedom consist of two scalar fields $\psi$ and
$\lambda$, the isotropic expansion rate $\dot\alpha$ and the anisotropic expansion rate $\dot\sigma$. Thus, our
solutions are specified by eight initial values for these parameters
and their time derivatives. The gravitational equations, however,
provide some relations between these parameters. Altogether, each
inflationary trajectory may  be specified by the values of six
parameters, $(g, \kappa;\ \psi_0, \dot\psi_0,
\lambda_0,\dot\lambda_0)$, here $0$ subscript indicates the  initial value.

In what follows we present the
results of the numerical analysis of the equations of motion
\eqref{lambda}, \eqref{const1} and the field equations corresponding to $\phi$ and $\lambda$, for
four sets of parameters  corresponding to  different positive
initial $\lambda$ values ($\lambda_0=1.05$, $\lambda_0=0.9$, $\lambda_0=0.1$ and
$\lambda_0=10$).
Note that $\kappa,\ \dot\alpha_0$, $\dot\sigma_0$ and $\psi_0$ are given in  units of $\mpl$.
 \vskip 2mm \noindent

\begin{figure}[h]
\includegraphics[angle=0, width=75mm, height=65mm]{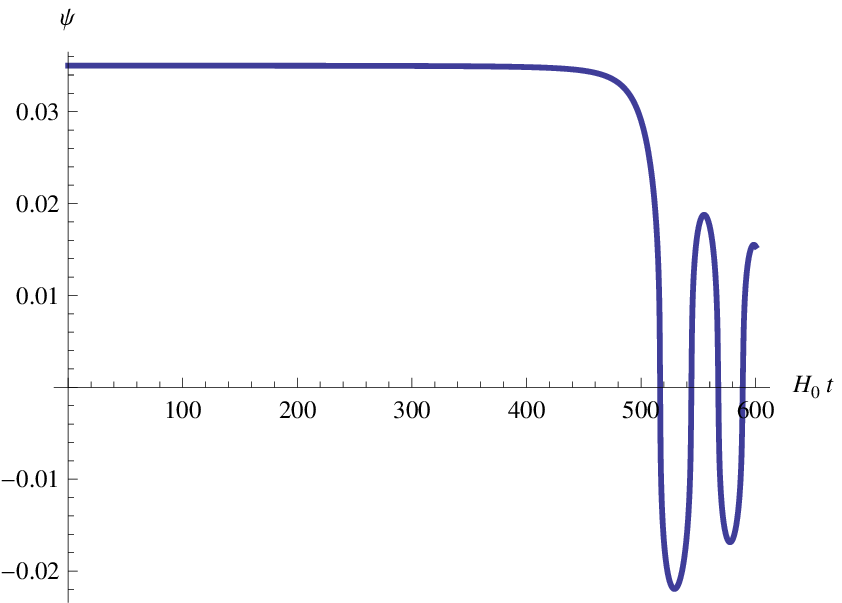}
\includegraphics[angle=0, width=75mm, height=65mm]{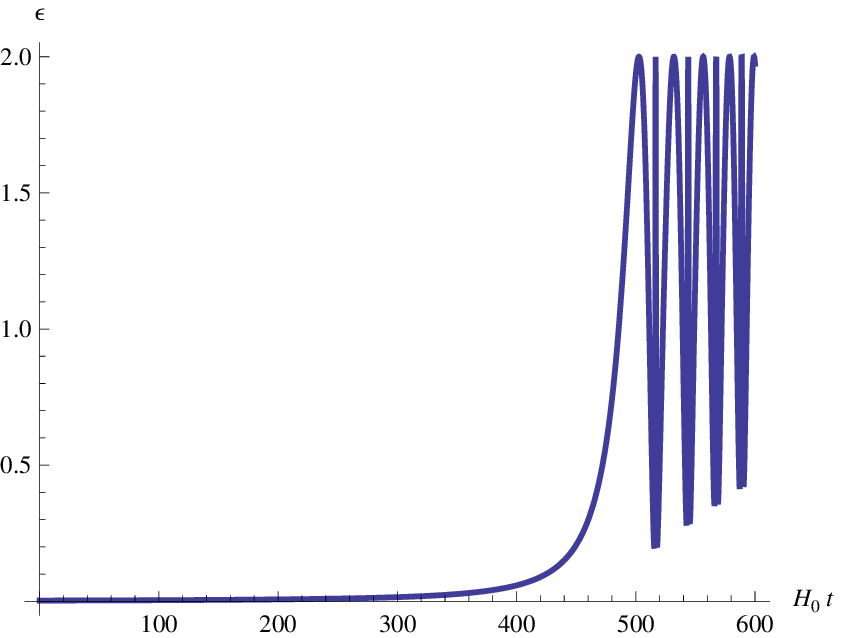}\\
\includegraphics[angle=0,width=75mm, height=65mm]{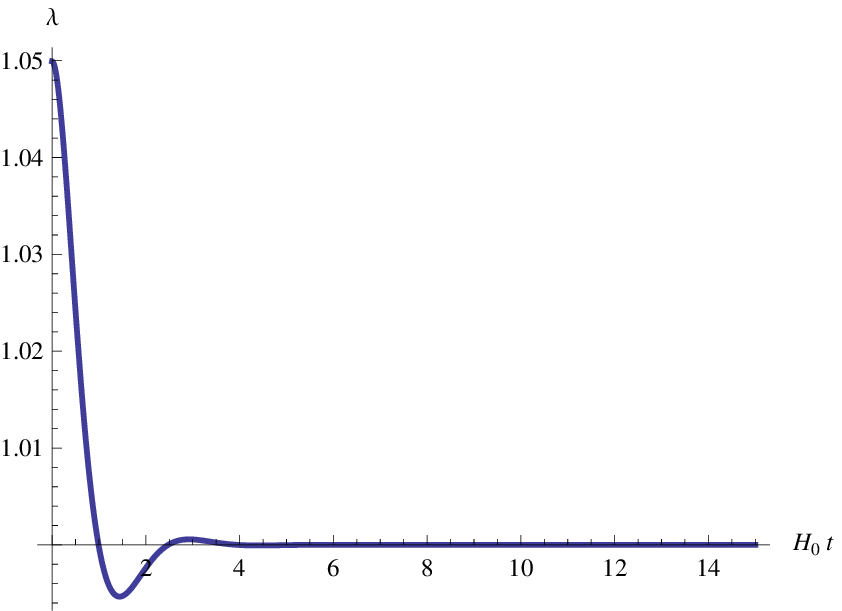}
\includegraphics[angle=0,width=75mm, height=65mm]{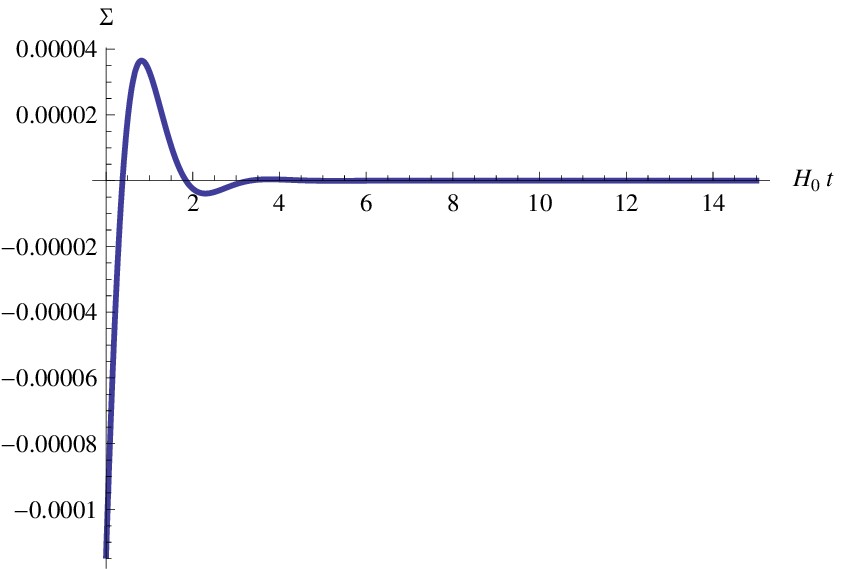}
\caption{The classical trajectory for ${\kappa = 1.733\times10^{14},\ g = 2.5\times10^{-3},\ \psi_0 = 0.034,\ \dot\psi_0 = -10^{-10},}$ ${\lambda_0 = 1.05,\ \dot\lambda_0 = 1.5\times10^{-10}}$. These values corresponds to a trajectory with $\dot\alpha_0=7.5\times
10^{-5},\  \epsilon_0=3\times 10^{-3}$ and $\dot\sigma_0=-8.6\times10^{-9}$. }\label{1.15-figures}
\end{figure}
\begin{figure}[h]
\includegraphics[angle=0, width=75mm, height=65mm]{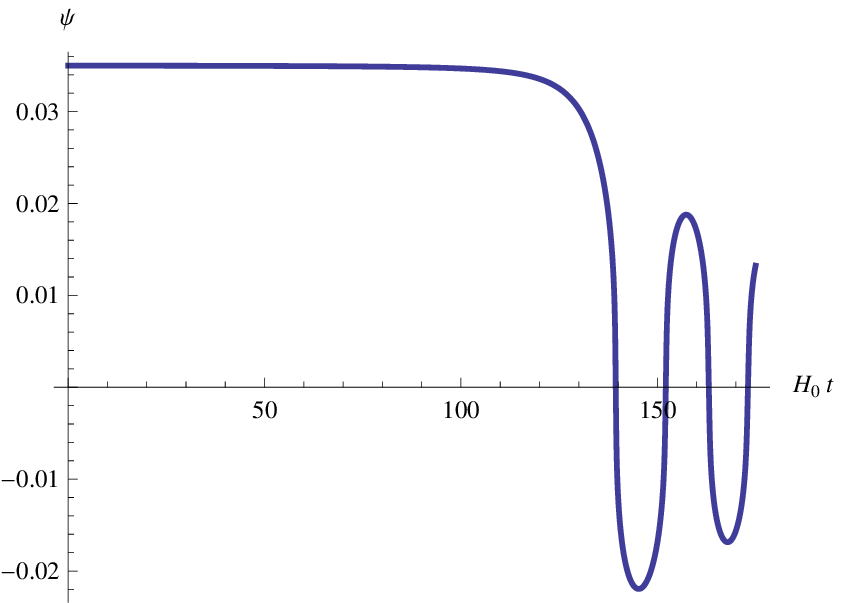}
\includegraphics[angle=0, width=75mm, height=65mm]{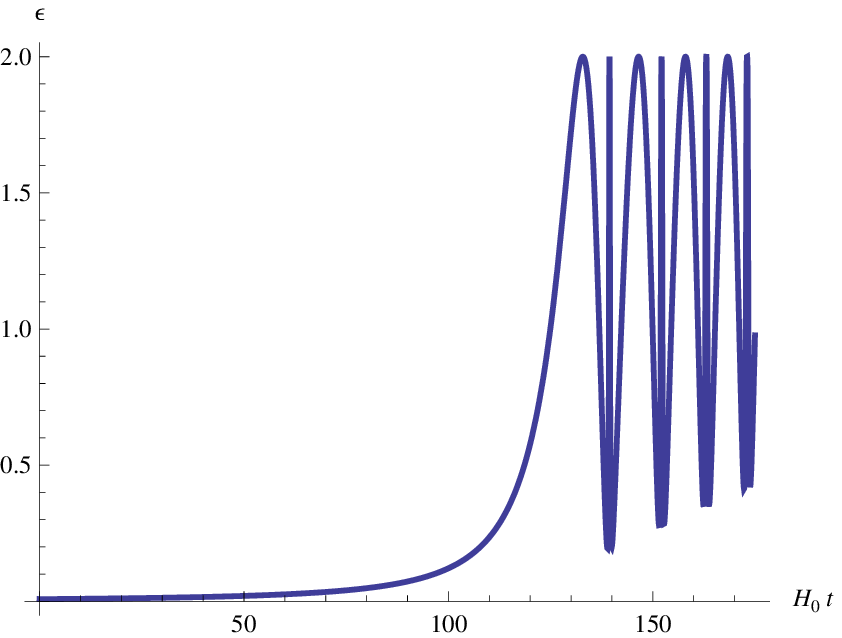}\\
\includegraphics[angle=0,width=75mm, height=65mm]{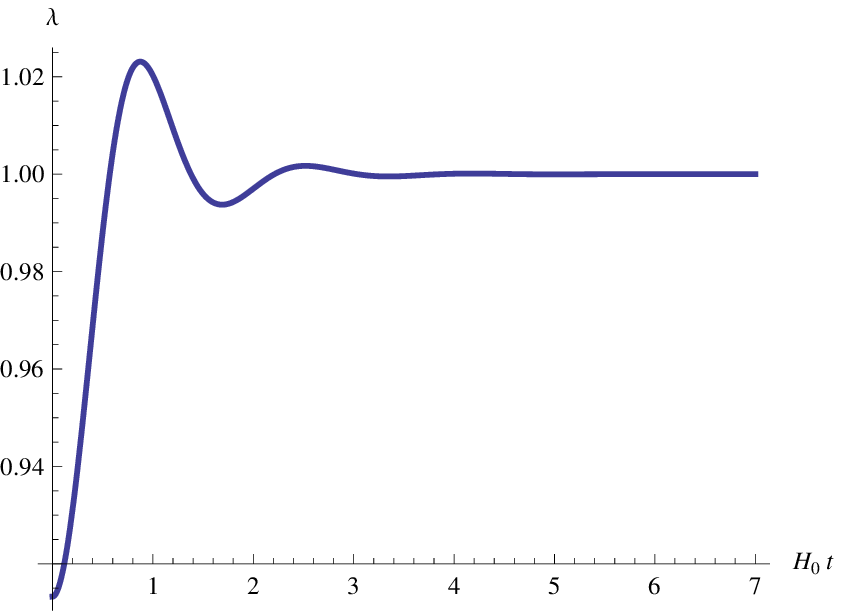}
\includegraphics[angle=0,width=75mm, height=65mm]{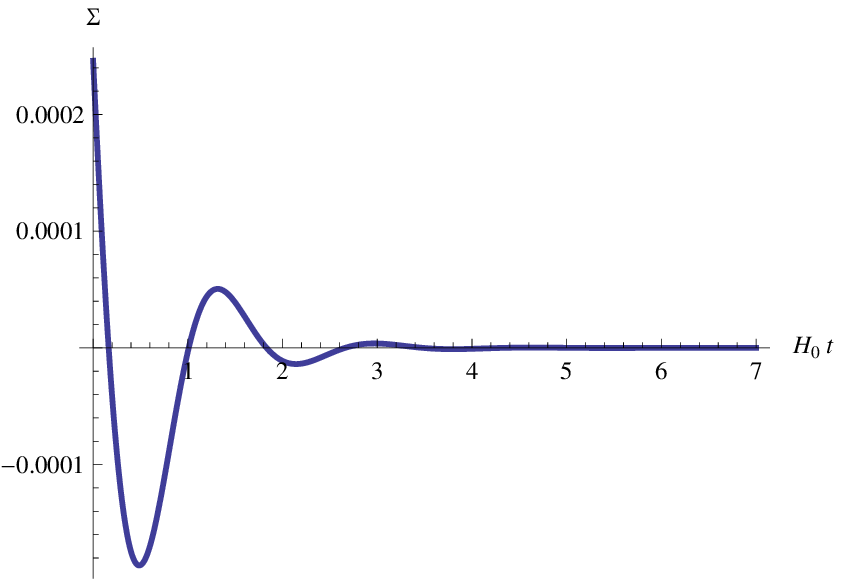}
\caption{The classical trajectory for ${\kappa=0.43\times10^{14},\ g = 5\times10^{-3},\ \psi_0 = 0.033,\ \dot\psi_0 = -0.8\times10^{-10},}$ ${\lambda_0 = 0.9,\ \dot\lambda_0 = 6\times10^{-10}}$. These values give a trajectory with $\dot\alpha_0=7\times 10^{-5},\  \epsilon_0=8\times 10^{-3}$, and $\dot\sigma_0=1.8\times10^{-8}$.}\label{0.75-figures}
\end{figure}

$\blacktriangleright$ \emph{\textbf{Discussion on diagrams in Fig.\ref{1.15-figures} and Fig.\ref{0.75-figures}}.}

Figures \ref{1.15-figures} and \ref{0.75-figures} show the classical trajectories of two systems initially close to the isotropic solution ($\lambda=1$). In other words, these systems start from $\lambda_0$ values in the vicinity of $\lambda=1$, which one of them has $\lambda_0>1$ and the other one has $\lambda_0<1$.

The top left figures in Figures \ref{1.15-figures} and \ref{0.75-figures} show classical trajectories of the field $\psi$ versus $\dot\alpha_0t$, while the top right figures show $\epsilon (t)$. As we see there is a period of quasi-de Sitter inflation, where $\psi,\ \epsilon$ remain almost constant and $\epsilon$ is very small. During this time, our initially anisotropic system mimics the behavior of the quasi-de Sitter inflation in the isotropic gauge-flation \cite{gauge-flation2}.
 Then, toward the end of the quasi-de Sitter inflation, $\epsilon$ grows and becomes equal to one (the top right figures), quasi-de Sitter inflation ends and $\psi$ suddenly falls off  and starts oscillating. As we see from the top right figures, the parameter $\epsilon$ has an upper limit which is equal to two. This is understandable recalling \eqref{aproxepsilon} and that $\rho_\kappa$ is positive definite.

The bottom left and right figures show evolutions of our two dimensionless variables $\lambda$ and $\Sigma$ during the first few e-folds respectively. As we see in the left bottom figures, $\lambda$ follows an underdamped oscillation and quickly approaches one . Furthermore, from the right bottom figures we learn that is quickly damped. After the quick damping of the anisotropies within a few e-folds, system almost follows an isotropic gauge-flation trajectory.

\begin{figure}[h]
\includegraphics[angle=0, width=75mm, height=65mm]{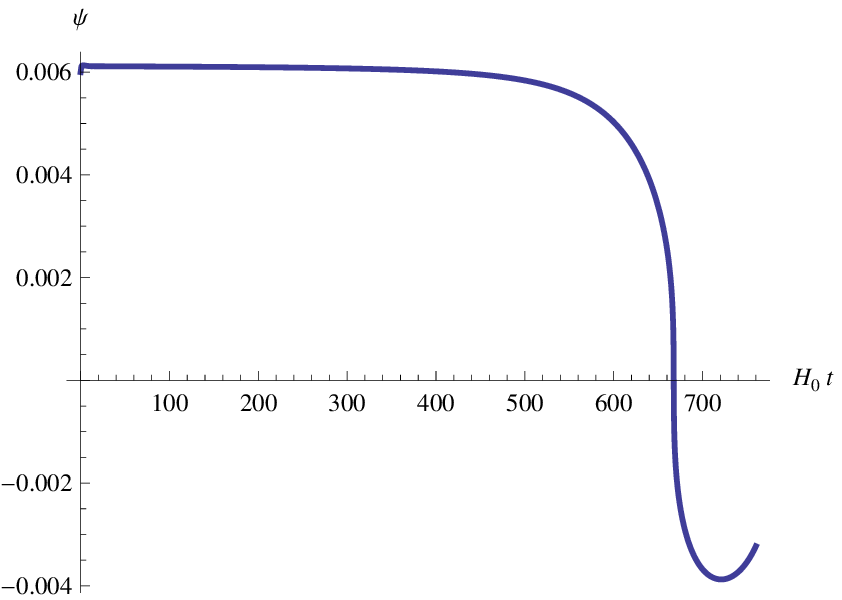}
\includegraphics[angle=0, width=75mm, height=65mm]{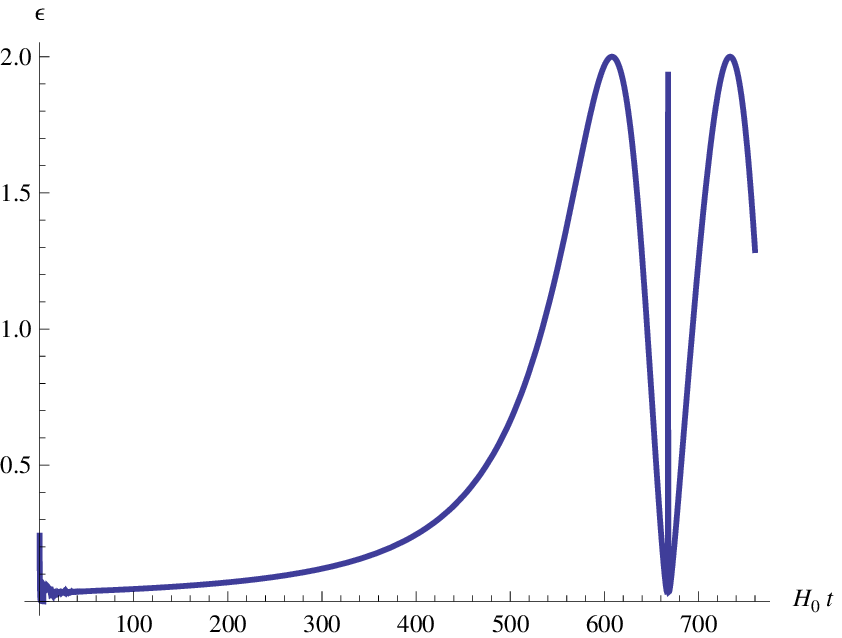}\\
\includegraphics[angle=0,width=75mm, height=65mm]{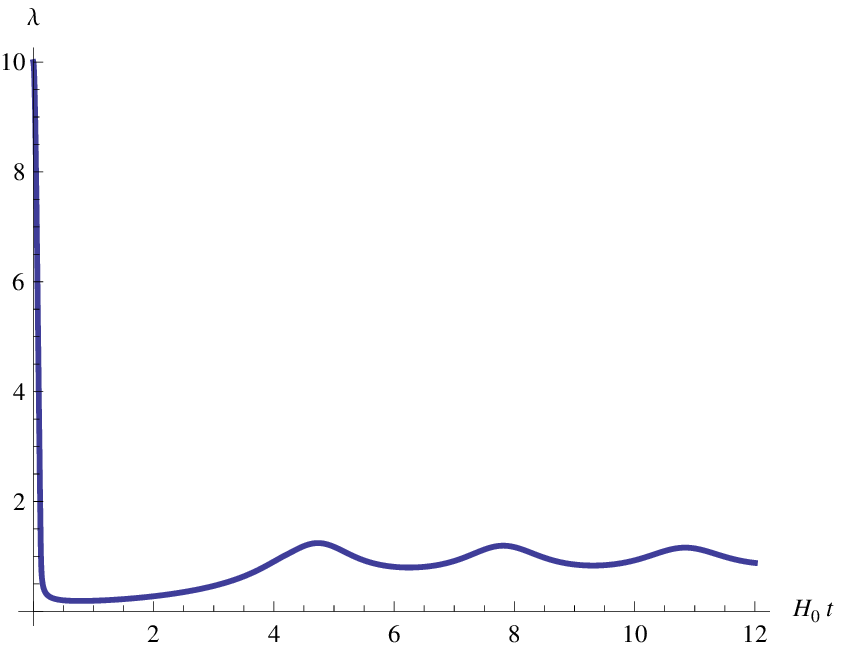}
\includegraphics[angle=0,width=75mm, height=65mm]{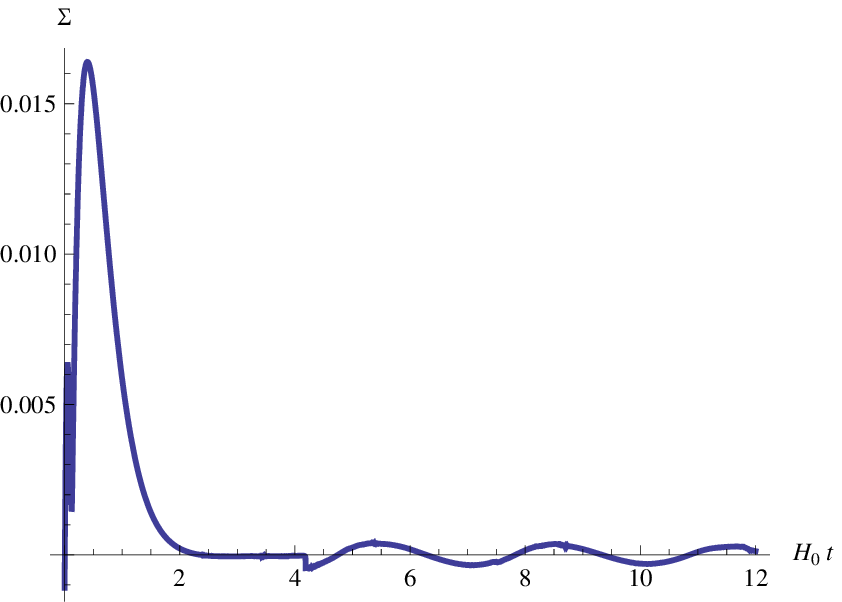}
\caption{The classical trajectory for ${\kappa = 3.77\times10^{15},\ g = 10^{-1},\ \psi_0 = 0.6\times 10^{-3},\ \dot\psi_0 = 10^{-10},}$
${\lambda_0 = 10,\ \dot\lambda_0 = -3.6}$. Here $\dot\lambda=-\dot\alpha\lambda$ which corresponds to $A_1=0$ case in \eqref{lam2} and as expected by our analytical calculations has a period in which $\Sigma$ is positive and increasing.
These values lead to a trajectory with $\dot\alpha_0=4\times10^{-4},\ \epsilon_0=0.24$ and $\dot\sigma_0=-5\times10^{-7}$. The initial value of $\epsilon$ is rather large, but with in one number of e-folds it decreases and reaches $10^{-2}$. Note that
value of $\epsilon$ at the point of maximum  $\Sigma$ is equal to $0.05$ ($\Sigma\simeq\frac13\epsilon$), almost saturating our upper bound for anisotrpy $\Sigma$.}\label{10-figures}
\end{figure}
\begin{figure}[h]
\includegraphics[angle=0, width=75mm, height=65mm]{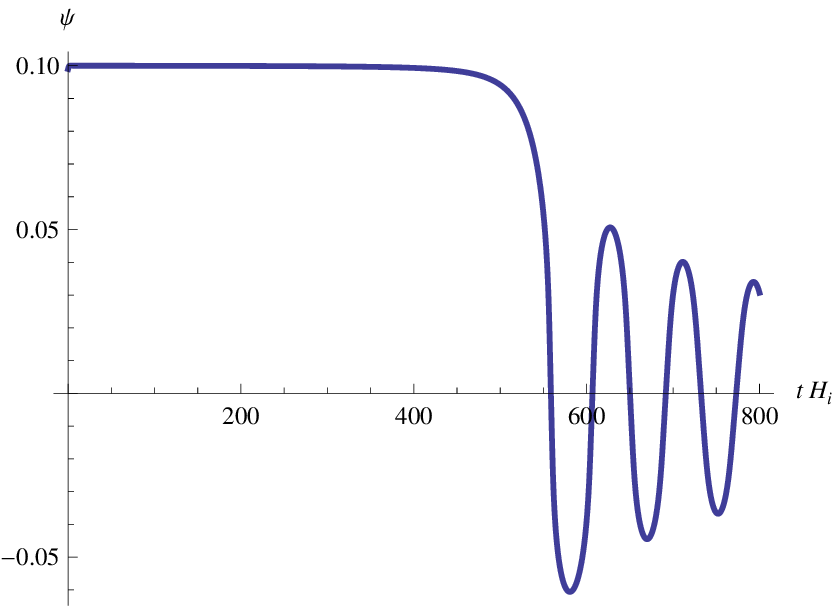}
\includegraphics[angle=0, width=75mm, height=65mm]{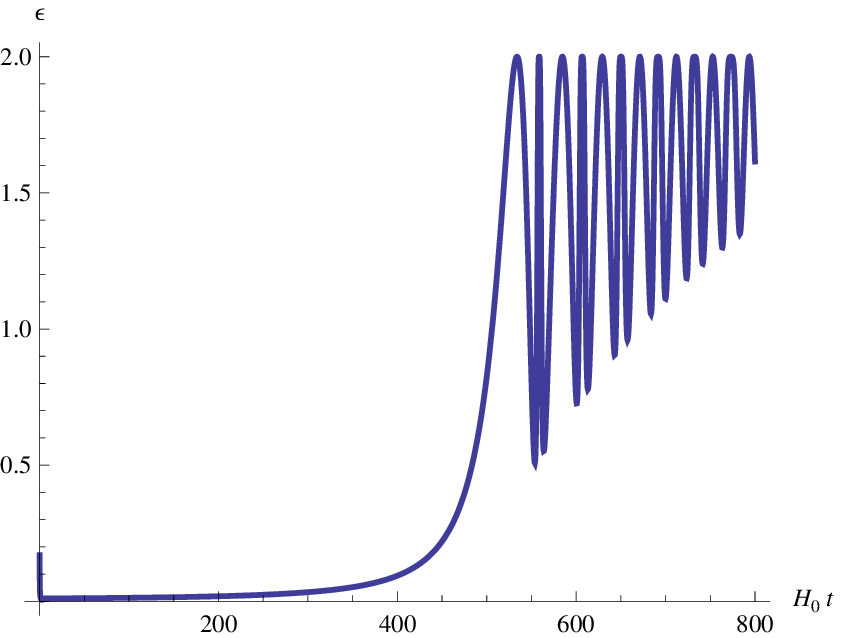}\\
\includegraphics[angle=0,width=75mm, height=65mm]{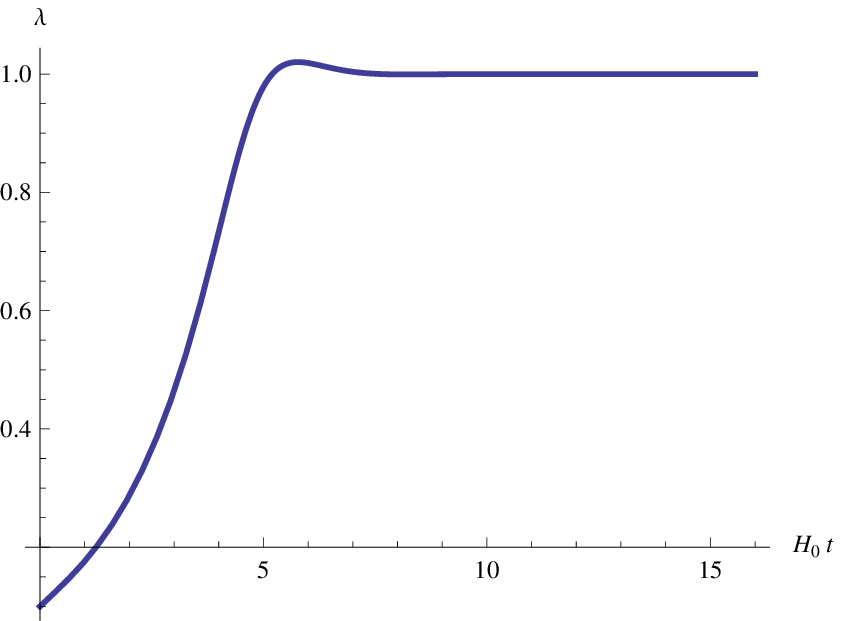}
\includegraphics[angle=0,width=75mm, height=65mm]{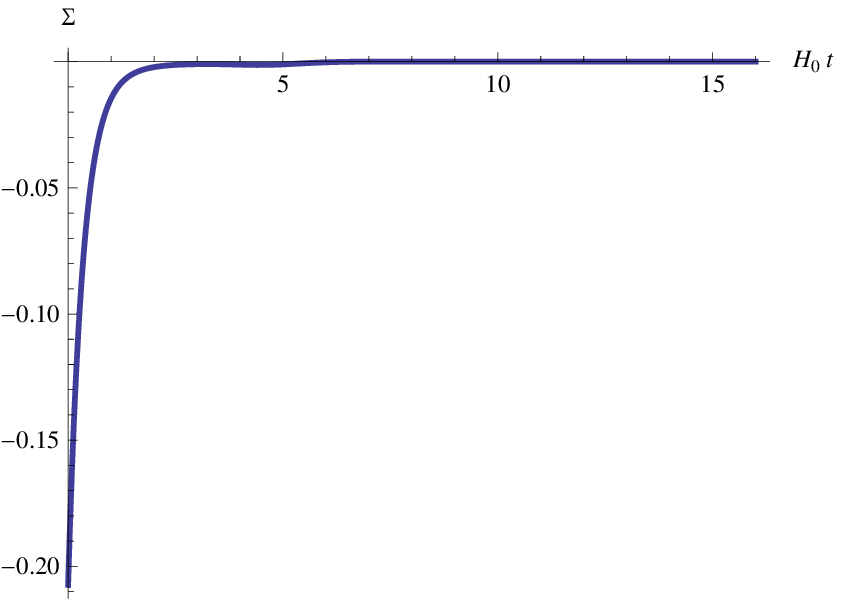}
\caption{The classical trajectory for ${\kappa = 1.99\times10^{10},\ g = 10^{-2},\ \psi_0 = 0.099,\ \dot\psi_0 = 10^{-10},}$ ${\lambda_0 = 0.1,\ \dot\lambda_0 = 1.288\times10^{-3}}$. Here $\dot\lambda=\dot\alpha\lambda$ which corresponds to the $A_2=0$ case in \eqref{lam1}. As has been predicted by the analytical calculations, $\Sigma$ is negative and monotonically damped.
These values give a trajectory with $\dot\alpha_0=4\times 10^{-3},\  \epsilon_0=0.16$ which is a rather large initial $\epsilon$ value. As we learn from the top right figure, $\epsilon$ decreases within a few number of $e$-folds to $0.5\times10^{-2}$. For these values $\dot\sigma_0=-0.8\times10^{-4}$.}\label{0.1-figures}
\end{figure}

$\blacktriangleright$ \emph{\textbf{Discussion on diagrams in Fig.\ref{10-figures} and Fig.\ref{0.1-figures}}.}

Figures \ref{10-figures} and \ref{0.1-figures} show the classical trajectories of two systems initially far from the isotropic solution ($\lambda^2=1$). These systems start from two positive large and small $\lambda$ values, $\lambda_0=10$ and $\lambda_0=0.1$.

Our analytical study reveals that although $\lambda$ evolves quickly toward the isotropic attractors $\lambda=\pm1$, there are two independent solutions for the evolution of $\Sigma$ in both of $\lambda^6\gg1$ and $\lambda^6\ll1$ limits.
In the first one, there is a short period of a few e-folds in which $\Sigma$ rapidly increases and then decreases in time. On the other hand in the second case, $\Sigma$ exponentially decreases.
Due to the similarities between the two behaviors of both limits and in order to make the long story short, we chose to plot a system of the first kind for $\lambda_0\gg1$ limit, and a system of the second kind for $\lambda_0\ll1$ limit. In Fig. \ref{10-figures}, a system with $\lambda=10$ and $\dot\lambda=-\dot\alpha\lambda$ is presented, which from the analytical results we expect to have a period of growing $\Sigma$. On the other hand, Fig. \ref{0.1-figures} shows a system with $\lambda=0.1$ and $\dot\lambda=\dot\alpha\lambda$, which we expect to have a monotonically decreasing $\Sigma$ term.

The top left figures in Figures \ref{10-figures} and \ref{0.1-figures} show classical trajectories of the field $\psi$ with respect to $\dot\alpha_0t$, while the top right figures indicate dynamics of $\epsilon$. As we see there is a period of quasi-de Sitter inflation, where $\psi$ remains almost constant and $\epsilon$ is almost
constant and very small. During this time, our initially anisotropic system mimics the behavior of the quasi-de Sitter inflation in the isotropic gauge-flation \cite{gauge-flation2}.
Then, toward the end of the quasi-de Sitter inflation, $\epsilon$ grows and becomes one (the top right figures), quasi-de Sitter inflation ends and $\psi$ suddenly falls off  and starts oscillating. As we see from the top right figures the parameter $\epsilon$ has an upper limit which is equal to two. This is understandable recalling \eqref{aproxepsilon} and that $\rho_\kappa$ is positive definite.

The bottom left and right figures show evolutions of our two dimensionless variables $\lambda$ and $\Sigma$ during the first few e-folds respectively.

In the left bottom figure of Fig. \ref{10-figures}, we see that $\lambda$ which started from $\lambda_0=10$ quickly decreases and gets close to one. The right bottom figure indicates that $\Sigma$, which is initially equal to $1.2\times10^{-3}$, shows a phase of rapid growth and saturates our upper bound $\Sigma \sim \epsilon$. More precisely, the peak value of $\Sigma$ is $\Sigma|_{t_{peak}}=0.016$ which is about $\frac{1}{3}\epsilon|_{t_{peak}}$.
After its sharp peak, $\Sigma$ decreases quickly and within  few e-folds becomes negligible. At that point, system mimics the behavior of isotropic inflation.

The left bottom figure of Fig. \ref{0.1-figures} shows $\lambda$, initially equal to $\lambda_0=0.1$, quickly evolves towards one. The right bottom figure indicates $\Sigma$, which is initially equal to $-2\times10^{-2}$, is exponentially damped and becomes negligible. As a result, after a few e-folds,  the system undergoes an essentially isotropic quasi-de Sitter inflation.

\section{Conclusion}\label{Section-conclusion}

We have studied nonlinear stability of the quasi-de Sitter gauge-flation with respect to the initial classical anisotropies.
Assuming a system which undergoes quasi-de Sitter inflation, in the sense that $\epsilon$, the dimensionless isotropic expansion rate time variation  \eqref{slow-roll-def}, is very small we examined the behavior of gauge-flation with respect to the anisotropies of the homogeneous background.
We showed both analytically and numerically that this system has two attractor solutions $\lambda\equiv\big(\frac{\psi_2}{\psi_1}\big)^\frac13=\pm1$ which regardless of the initial values of $\lambda$, all the solutions converge to them within a few e-folds. Here $\psi_1$ and $\psi_2$ are our field values.
These two attractor branches, which are physically identical due to parity and charge conjugation invariance of our gauge-flation action \eqref{The-model}, correspond to the isotropic quasi-de Sitter solutions. Thus, gauge-flation is globally stable with respect to the initial anisotropies. We note that this is not a trivial result because in our system we have turned on vector gauge fields in the background and that the vector perturbations have a non-trivial source \cite{gauge-flation2}.

We showed that this attractor behavior happens for generic initial values of anisotropy $\lambda$ and the convergence to the isotropic attractor point happens very fast, within a few e-folds. This behavior may be contrasted with the anisotropic inflation model discussed in \cite{Jiro-1, Jiro-observational-imprint}, where the model exhibits
an anisotropic attractor with non-zero, but nevertheless small $\Sigma\sim \epsilon$ at the end of inflation.

Due to the fast damping of anisotropies in our model and that they do not last up to five to ten e-folds before the end of inflation, gauge-flation predicts no detectable features of statistical anisotropy in the CMB temperature-temperature correlation function. In this respect gauge-flation is similar to the standard scalar-driven inflationary models and is unlike the models discussed in \cite{Jiro-1,{Jiro-nature},Jiro-observational-imprint}.

Furthermore, we found an upper bound on the value of $|\Sigma|=\frac{|\dot\sigma |}{\dot\alpha}\sim {\epsilon}$. Our numerical analysis reveals that
in the extreme limits of $\lambda^6\gg1$ and $\lambda^6\ll1$, there is the possibility
to saturate our upper bound for a very short lapse of time. However, after reaching this maximum value, $\Sigma$ is exponentially damped and soon becomes negligible.

Here we focused mainly on the Bianchi type-I homogenous, anisotropic background metric. However, we expect our result, the global stability of the quasi-de Sitter gauge-flation against the initial anisotropies and its isotropic attractor fixed point, extend over the other Bianchi type models.
(Similar analysis for the case of power-law anisotropic inflation model \cite{Jiro-power-law} has been carried out in \cite{Bianchi-II-III}.) This expectation is based on the fact that the global stability observed here is a result of the $\rho_k$ dominance (\emph{cf.} discussions in the opening of section \ref{Section-II}) and that this term can be written in terms of the single ``effective inflaton'' field $\phi$ (\emph{cf}. \eqref{phi-limit}, \eqref{rho0-rho1}). These  features still hold for general Bianchi backgrounds. We will expand on this argument in upcoming work \cite{extended Wald theorem}.

Our analytic and numerical computations shows that for very large and small initial values of $\lambda$ there is a region where anisotropy grows exponentially for a very short period, before getting exponentially damped to the isotropic fixed point. Although in these cases the system does not follow the strict dynamics indicated by the Wald's cosmic no-hair theorem \cite{Wald theorem} for the very short period of time, the anisotropies are indeed damped within few Hubble times, in accord with the cosmic no-hair conjecture. In more general viewpoint, one can show that in general inflationary models do not strictly obey dominant energy condition assumption of cosmic no-hair theorem (\emph{cf}. discussions in the second paragraph of the introduction) and hence in principle there is a possibility of violating this theorem. This possibility can be realized and can be more pronounced in the inflationary models involving vector (gauge) fields. In \cite{extended Wald theorem} we will extend cosmic no-hair theorem such that it is also applicable to the cases like our gauge-flation and the anisotropic inflation model \cite{Jiro-1}.

\section*{Acknowledgments}
M.M.Sh-J would like to thank Masud Chaichian and Anca Tureanu for  discussions.
JS is supported by  the
Grant-in-Aid for  Scientific Research Fund of the Ministry of
Education, Science and Culture of Japan No.22540274, the Grant-in-Aid
for Scientific Research (A) (No.21244033, No.22244030), the
Grant-in-Aid for  Scientific Research on Innovative Area No.21111006,
and JSPS under Japan-Russia Research Cooperative Program.



\begin{thebibliography}{99}

\bibitem{gauge-flation1}
  A.~Maleknejad, M.~M.~Sheikh-Jabbari,
  ``Gauge-flation: Inflation From Non-Abelian Gauge Fields,''[arXiv:1102.1513 [hep-ph]].

\bibitem{gauge-flation2}
A. Maleknejad, M.M. Sheikh-Jabbari, ``Non-Abelian Gauge Field Inflation,''
Phys. Rev. D 84, 043515 (2011),
[arXiv:1102.1513 [hep-ph]].



\bibitem{WMAP7-data}
  E.~Komatsu {\it et al.} [ WMAP Collaboration ],
  ``Seven-Year Wilkinson Microwave Anisotropy Probe (WMAP) Observations: Cosmological Interpretation,''
[arXiv:1001.4538 [astro-ph.CO]].



\bibitem{Inflation-texts}

V. Mukhanov, ``Physical Foundations of Cosmology,'' Cambrdige Uni. Press (2005);

D. Lyth and A. Liddle, ``Primordial Density Perturbations,'' Cambridge Uni. Press (2009).

  B.~A.~Bassett, S.~Tsujikawa and D.~Wands,
 ``Inflation dynamics and reheating,''
  Rev.\ Mod.\ Phys.\  {\bf 78}, 537 (2006),
  [arXiv:astro-ph/0507632];

\bibitem{Ellis:1998ct}
  G.~F.~R.~Ellis, H.~van Elst,
  ``Cosmological models: Cargese lectures 1998,''
  NATO Adv.\ Study Inst.\ Ser.\ C.\ Math.\ Phys.\ Sci.\  {\bf 541}, 1-116 (1999).
  [gr-qc/9812046].

\bibitem{Wald theorem}
R. Wald, ``Asymptotic behavior of homogeneous cosmological models in the presence of a positive cosmological constant,''  Phys.\ Rev.\  {\bf D28}, 2118 (1983).


\bibitem{Sahni}
  I.~Moss, V.~Sahni,
``Anisotropy In The Chaotic Inflationary Universe,''
  Phys.\ Lett.\  {\bf B178}, 159 (1986).



\bibitem{Ford:1989me}
  L.~H.~Ford, ``Inflation Driven By A Vector Field,''
  Phys.\ Rev.\  {\bf D40}, 967 (1989).

\bibitem{vector-inflation}
A. Golovnev, V. Mukhanov and V. Vanchurin, ``Vector Inflation,''JCAP 0806, 009 (2008), [arXiv:0802.2068[astro-ph]].

\bibitem{vector-inflation-loophole}

B.~Himmetoglu, C.~R.~Contaldi, M.~Peloso,
  ``Instability of anisotropic cosmological solutions supported by vector fields,''
  Phys.\ Rev.\ Lett.\  {\bf 102}, 111301 (2009), [arXiv:0809.2779 [astro-ph]];
``Instability of the ACW model, and problems with massive vectors during inflation,''
  Phys.\ Rev.\  {\bf D79}, 063517 (2009).
  [arXiv:0812.1231 [astro-ph]].



\bibitem{Jiro-1}
 M.~-a.~Watanabe, S.~Kanno, J.~Soda,
  ``Inflationary Universe with Anisotropic Hair,''
  Phys.\ Rev.\ Lett.\  {\bf 102}, 191302 (2009),
  [arXiv:0902.2833 [hep-th]].

\bibitem{Jiro-nature}
M.~-a.~Watanabe, S.~Kanno, J.~Soda,
  ``The Nature of Primordial Fluctuations from Anisotropic Inflation,''
  Prog.\ Theor.\ Phys.\  {\bf 123}, 1041-1068 (2010).
  [arXiv:1003.0056 [astro-ph.CO]];\\
  T.~R.~Dulaney, M.~I.~Gresham,
  ``Primordial Power Spectra from Anisotropic Inflation,''
  Phys.\ Rev.\  {\bf D81}, 103532 (2010).
  [arXiv:1001.2301 [astro-ph.CO]];\\
  A.~E.~Gumrukcuoglu, B.~Himmetoglu, M.~Peloso,
  ``Scalar-Scalar, Scalar-Tensor, and Tensor-Tensor Correlators from Anisotropic Inflation,''
  Phys.\ Rev.\  {\bf D81}, 063528 (2010),
  [arXiv:1001.4088 [astro-ph.CO]].

\bibitem{Jiro-observational-imprint}
  M.~-a.~Watanabe, S.~Kanno, J.~Soda,
  ``Imprints of anisotropic inflation on the cosmic microwave background,''
  Mon.\ Not.\ Roy.\ Astron.\ Soc.\  {\bf 412}, L83-L87 (2011),
  [arXiv:1011.3604 [astro-ph.CO]].



\bibitem{Jiro-nonAbelian}
  K.~Murata, J.~Soda,
  ``Anisotropic Inflation with Non-Abelian Gauge Kinetic Function,''
  JCAP {\bf 1106}, 037 (2011),
  [arXiv:1103.6164 [hep-th]].

\bibitem{Jiro-power-law}
  S.~Kanno, J.~Soda, M.~-a.~Watanabe,
``Anisotropic Power-law Inflation,''
  JCAP {\bf 1012}, 024 (2010),
  [arXiv:1010.5307 [hep-th]].

\bibitem{anisotropic-others}
R.~Emami, H.~Firouzjahi, S.~M.~Sadegh Movahed, M.~Zarei,
``Anisotropic Inflation from Charged Scalar Fields,'' JCAP {\bf 1102}, 005 (2011), [arXiv:1010.5495 [astro-ph.CO]];

K.~Dimopoulos, J.~M.~Wagstaff,
``Particle Production of Vector Fields: Scale Invariance is Attractive,''
Phys.\ Rev.\ {\bf D83}, 023523 (2011), [arXiv:1011.2517 [hep-ph]].

\bibitem{non-abelian-others}
N.~Bartolo, E.~Dimastrogiovanni, S.~Matarrese, A.~Riotto,
``Anisotropic bispectrum of curvature perturbations from primordial non-Abelian vector fields,''
JCAP {\bf 0910}, 015 (2009),  [arXiv:0906.4944 [astro-ph.CO]];
``Anisotropic Trispectrum of Curvature Perturbations Induced by Primordial Non-Abelian Vector Fields,''
  JCAP {\bf 0911}, 028 (2009),  [arXiv:0909.5621 [astro-ph.CO]]; ``Non-Gaussianity and statistical anisotropy from vector field populated inflationary models,''
  Adv.\ Astron.\  {\bf 2010}, 752670 (2010),
  [arXiv:1001.4049 [astro-ph.CO]].




\bibitem{Bianchi-II-III}
  S.~Hervik, D.~F.~Mota, M.~Thorsrud,
  ``Inflation with stable anisotropic hair: is it cosmologically viable?,''
  [arXiv:1109.3456 [gr-qc]].



\bibitem{extended Wald theorem}
A. Maleknejad, M.M. Sheikh-Jabbari and Jiro Soda, \emph{Work in progress}.


\end{thebibliography}
\end{document}